\documentclass{article}

\PassOptionsToPackage{numbers, compress}{natbib}
\usepackage[final]{neurips_2025_ml4ps}

\usepackage[utf8]{inputenc} 
\usepackage[T1]{fontenc}    
\usepackage{url}            
\usepackage{booktabs}       
\usepackage{amsfonts}       
\usepackage{nicefrac}       
\usepackage{microtype}      
\usepackage{subcaption} 
\usepackage{booktabs,adjustbox}

\usepackage{graphicx}
\usepackage{dcolumn}
\usepackage{bm}
\usepackage{tensor}
\usepackage{amsmath}
\usepackage{mathrsfs}
\usepackage{amsthm,amsfonts,amssymb}
\usepackage{aas_macros}
\usepackage{CJKutf8}
\usepackage{orcidlink}
\usepackage{soul}
\usepackage{enumitem}


\usepackage{fix-cm}
\usepackage{amsmath}
\usepackage{soul}
\usepackage{xcolor}
\usepackage{hyperref}
\usepackage{CJKutf8}
\usepackage{appendix}

\usepackage{cleveref} 

\usepackage{graphicx}
\usepackage{duckuments}
\usepackage{tikzducks}
\usepackage{orcidlink}
\renewcommand{\arraystretch}{1.1}
\DeclareMathOperator{\sgn}{sgn}

\begin{document}
\begin{CJK*}{UTF8}{gbsn}
    \title{From Black Hole to Galaxy: Neural Operator Framework for Accretion and Feedback Dynamics}
\author{
Nihaal Bhojwani$^{1,2*}$, Chuwei Wang$^{2*}$, Hai-Yang Wang$^{3*}$, Chang Sun$^{4}$,\\ 
\textbf{Elias R. Most$^{3}$, Anima Anandkumar$^{2}$}\\[3pt]
$^{1}$Department of Computer, Mathematical, and Natural Sciences, University of Maryland,\\
$^{2}$Department of Computing and Mathematical Sciences, California Institute of Technology,\\
$^{3}$TAPIR \& Walter Burke Institute for Theoretical Physics, California Institute of Technology\\
$^{4}$Department of Physics, California Institute of Technology\\[3pt]
\texttt{nbhojwan@umd.edu, \{chuweiw, haiyangw, chsun, emost, anima\}@caltech.edu}
}

\maketitle

\footnotetext[1]{\hspace{-0.5em}$^{*}$Equal contribution. In alphabetical order.}

\vspace{-0.6em}
    \begin{abstract}
    \vspace{-0.6em}
        Modeling how supermassive black holes co-evolve with their host galaxies is notoriously hard because the relevant physics spans nine orders of magnitude in scale—from milliparsecs to megaparsecs—making end-to-end first-principles simulation infeasible. 
        To characterize the feedback from the small scales, existing methods employ a static subgrid scheme or one based on theoretical guesses, which usually struggle to capture the time variability and derive physically faithful results.
        Neural operators are a class of machine learning models that achieve significant speed-up in simulating complex dynamics. We introduce a neural-operator–based \textit{``subgrid black hole''} that learns the small-scale local dynamics and embeds it within the direct multi-level simulations. 
        Trained on small-domain (general relativistic) magnetohydrodynamic data, the model predicts the unresolved dynamics needed to supply boundary conditions and fluxes at coarser levels across timesteps, enabling stable long-horizon rollouts without hand-crafted closures. Thanks to the great speedup in fine-scale evolution, our approach for the first time captures intrinsic variability in accretion-driven feedback, allowing dynamic coupling between the central black hole and galaxy-scale gas. This work reframes subgrid modeling in computational astrophysics with scale separation and provides a scalable path toward data-driven closures for a broad class of systems with central accretors.
    \end{abstract}
\vspace{-1em}
\end{CJK*}

\section{Introduction}
\label{sec:intro}
\vspace{-0.2em}
\textbf{Background.} Supermassive black holes (SMBHs) co-evolve with their host galaxies through a two-way feeding-feedback loop that operates across nine orders of magnitude in scale~\cite{Kormendy:2013dxa}. 
Accretion flows near the event horizon (milliparsec scales) are governed by general relativistic magnetohydrodynamics (GRMHD) and launch relativistic jets and winds. 
These outflows propagate from AU to megaparsec scales, well beyond the host galaxy, depositing energy and momentum into the interstellar and intergalactic media~\cite{Blandford:2019ARA&A..57..467B,McNamara:2007ww}. 
The resulting feedback regulates star formation and galaxy growth, while galaxy-scale gas dynamics in turn modulate the inflow feeding the central black hole. 

\begin{figure}[htbp]
    \centering
    \includegraphics[width=0.95\textwidth]{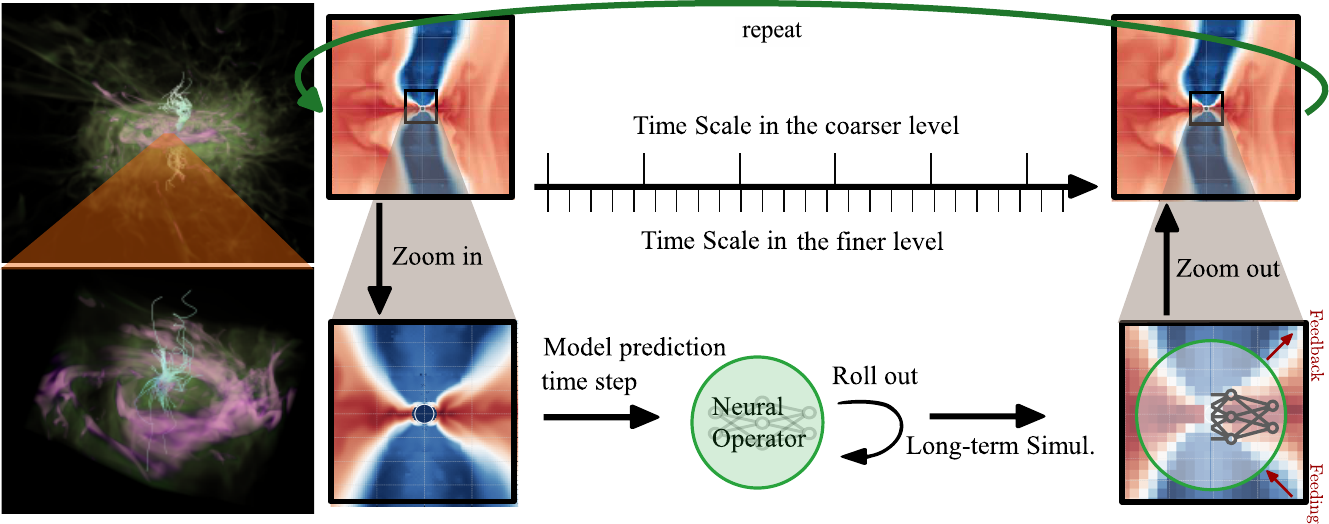}
    \caption{Illustration of our method in a two-level setting. The neural operator efficiently replaces the evolution on small scales (zoom in), enabling proper feedback on large scale (zoom out).
    }\label{fig:problem}
    \vspace{-2em}
\end{figure}

\textbf{Challenge.}
This multiscale coupling makes end-to-end simulation computationally intractable: accurately resolving accretion flows demands timesteps set by the gravitational radius ($r_g \sim \mathrm{mpc}$ for M87), while capturing galaxy-scale feedback requires following dynamics over $\sim 10^9$ times larger spatial and temporal scales. 
Bridging these domains requires passing information across many orders of magnitude in space and time without washing out variability or violating conservation.

A considerable amount of efforts adopting various techniques has been made to bridge different scales, including direct simulations under reduced scale separation~\cite{Lalakos:2023ean,Lalakos:2022qhl,S:2023gta,Lalakos:2025msz},
nested mesh “zoom-in’’ approaches in Eulerian codes~\cite{Guo2023,Guo2024},
“Lagrangian hyper-refinement’’ in Larangian codes~\cite{Hopkins:2023ipv,Hopkins2023},
remapping between domains~\cite{Kaaz:2024jxl}, and iterative “multi-zone’’~\cite{Cho:2023wqr,Cho:2024wsp,Cho:2025lzq} or “cyclic-zoom’’~\cite{Guo:2025sjb} schemes that repeatedly refine/de-refine around the SMBH.
Among these, multi-zone and cyclic-zoom methods propagate small-scale feedback to larger domains and evolve to consistency, but can still struggle with temporal variability and with specifying a physically faithful inner boundary (here we name this by \emph{subgrid black hole}) across levels. 
The inner boundary adopted this way could feed erroneous information (e.g., in both methods, the relativistic jet cannot be turned on/off when crossing the boundary) to the active simulation domain, consequently alter the final results.

\textbf{Neural Operator.}
Neural operators (NOs) represent a recent advance in machine learning for partial differential equations, learning mappings between infinite-dimensional function spaces rather than approximating individual functions~\cite{li2020fourier,kovachki2023neural}. This approach has demonstrated substantial computational acceleration and shown promise as surrogate models across various applications~\cite{plasma-surrogate:2024NucFu..64e6025G,pathak2022fourcastnet,wen2023real, tolooshams2025equireg,Wang:2024arXiv240805177W,wang2025accelerating}. Recent studies have begun exploring NOs for coarse graining or closure modeling problems~\cite{Wang:2024arXiv240805177W}. 
Astrophysical systems present unique challenges. Realistic galaxy formation involves large-scale, strongly coupled dynamics with central singularities at supermassive black holes, requiring stable long-term integration over cosmological timescales. The extreme dynamic range, combined with the inherent scarcity of training data in astrophysical simulations, creates a particularly demanding test case for NO or other machine learning methods. To the best of our knowledge, despite some preliminary attempts in simplified systems or transient-state predictions \cite{ohana2024well,duarte2022black,poletti2025modeling,kacmaz2025resolving}, no previous work has demonstrated NO performance in such complex, multi-scale systems with limited training data.

\textbf{Contribution.}
We introduce a neural-operator-based \emph{``subgrid black hole''} coupled to a direct multi-level solver, replacing hand-crafted closure rules with a data-driven model that captures realistic variability in accretion and feedback (see \cref{fig:problem} for a schematic illustration of our method):

\emph{(1) Operator-learning subgrid.} We train a neural operator to approximate the small-scale (GR)MHD time-evolution semigroup. The learned model supplies boundary conditions and fluxes to the next coarser level, enabling long-horizon rollouts without assuming steady or time-averaged injection.
\emph{(2) Two-way multiscale coupling.} Embedded within a multi-level framework, the learned subgrid responds to evolving large-scale conditions and, in turn, drives feedback that propagates outward—preserving variability critical to galaxy–SMBH co-evolution.
\emph{(3) General applicability.} The demonstrated approach is broadly applicable as a subgrid model for systems with a central accretor (e.g., SMBHs and neutron stars), independent of problem setups and generalizable to different codes.

This work reframes subgrid modeling in computational astrophysics as \emph{operator learning}: rather than prescribing fixed closures, we learn the small-scale dynamics to provide dynamically-updated boundary conditions for larger domains.
{This advance in subgrid modeling fills up the missing piece for the black hole accretion and feedback problem. 
More importantly, this could revolutionize the modeling of black hole feedback in cosmological simulations such as FIRE~\cite{Hopkins-fire3:2023MNRAS.519.3154H} and IllustrisTNG~\cite{Nelson:2018uso}.}

\textbf{Related Works.} Analytical subgrid prescriptions remain the standard tool in large-scale such as cosmological simulations (e.g., supernova feedback, cold-gas processes, and black-hole growth~\cite{Hopkins:2017ijo,Butsky:2024MNRAS.535.1672B}). 
Machine-learning–based subgrid/surrogate models are only beginning to appear, with early results for supernova feedback~\cite{Hirashima:2023MNRAS.526.4054H,Hirashima:2025ApJ...987...86H} and for 2D black-hole accretion surrogates~\cite{Duarte:2022MNRAS.512.5848D} trained on Newtonian hydrodynamical simulation of steadily accreting torus. Our work advances this line by learning an operator-level ``subgrid black hole'' compatible with multi-level MHD/GRMHD.

\begin{figure}[t]
    \centering
    \includegraphics[width=0.95\linewidth]{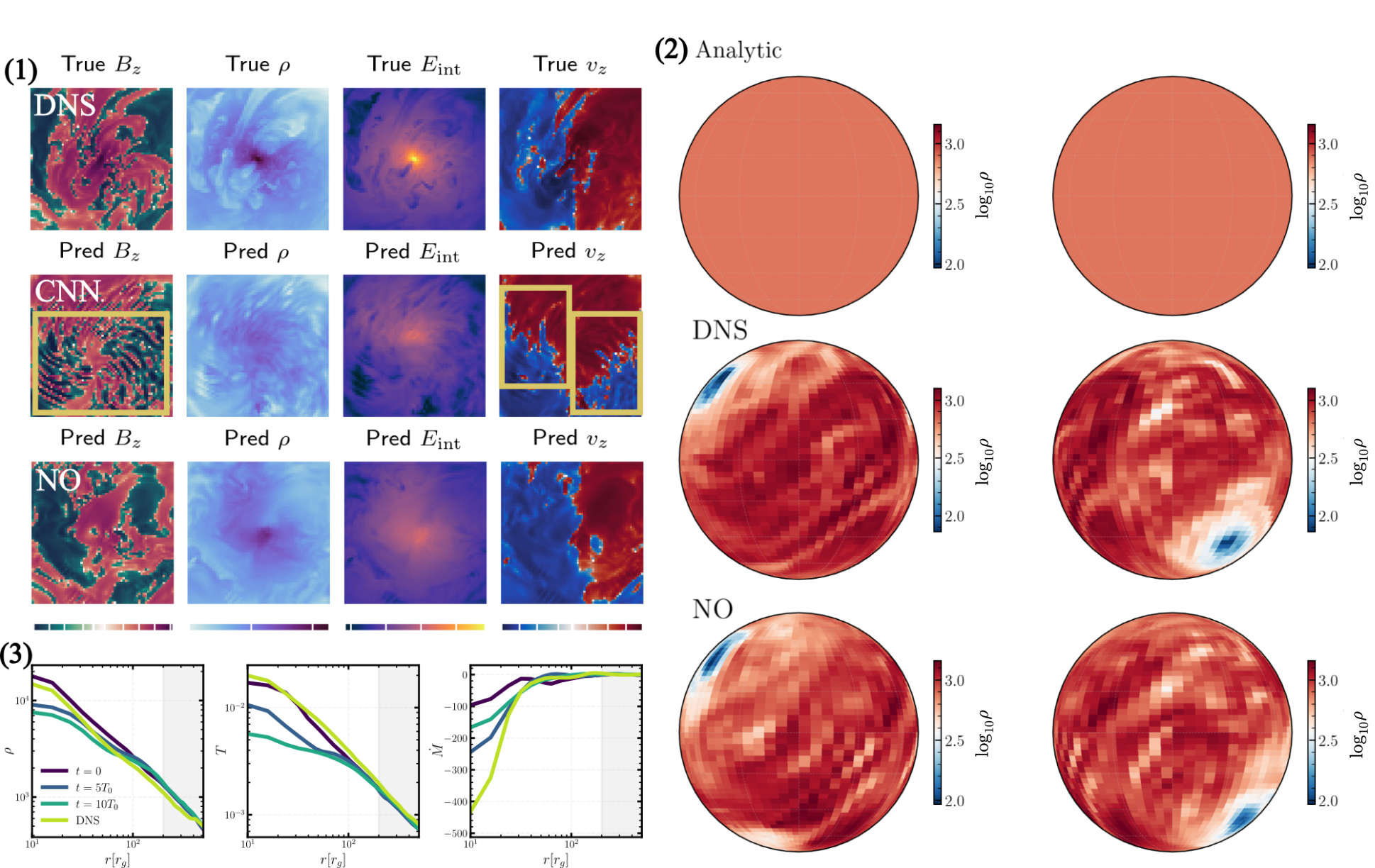}
    \caption{\textbf{MHD outputs.}
        Simulated (`DNS'), CNN-predicted, and NO-predicted magnetic field component \(B_z\), density \(\rho\), internal energy \(E_{int}\), and velocity component \(v_z\) from the MHD simulation on a horizontal slice through the domain center  (after 50 steps). Unphysical artifacts are marked.
        (2) Comparison of the mass density distributions on a 2D sphere, between analytically prescribed, directed simulated, and NO-rollout (after 10 steps).
        (3) The radial profiles of predicted quantities: mass density $\rho$, temperature $T$, and total mass flux $\dot{M}$ (\textit{Left to Right}). Profile from simulation after 50 steps is also shown. 
        The grey-shaded region shows the part that will be coupled to the direct simulation~(see Appendix \ref{apdx: insert back}).
        }
    \label{fig:mhd-outputs}
    \vspace{-2.0em}
\end{figure}

\vspace{-0.9em}
\section{Method}\label{sec:method}
\vspace{-0.8em}

\textbf{Method Framework.}
We consider the (general relativistic) magnetohydrodynamics (GRMHD) equations on a bounded spatial domain \(\Omega \subset \mathbb{R}^3\) over a finite time interval. See Appx. \ref{sec:method-simulation} for governing equations and initial conditions.
We introduce a two-level neural-operator--direct-numerical-simulation framework for the accretion--feedback problem with a \emph{``subgrid black hole''} model. Our conventions are: (i) \textbf{domain decomposition}: a coarse level (with domain size $(n_LL)^3$) and a fine level (with domain size $(L)^3$) are defined, $n_L$ is a large number which we choose to be $6$; (ii) \textbf{domain treatment}: within affordable simulations, the coarse level is evolved via DNS, while the fine level is unresolved and represented by a subgrid model; (iii) \textbf{domain coupling}: resolving the fine level is ultimately necessary to address the multiscale problem discussed in Sec.~\ref{sec:intro}. The framework is summarized in Fig.~\ref{fig:problem}. 

Directly simulating the fine and coarse levels together would shorten the timestep, thus increasing the cost, by a factor of at least \(n_L^2\). Instead, we first simulate the fine level over \(t \in [0, T]\), producing \(N_{\mathrm{data}} = 300\) snapshots (saved every $\Delta T:=T/N_{data}$) as the training set. We then train a local neural operator~\cite{liu-schiaffini2024neural} using the fine level simulation data. This neural operator is coupled to the coarse level, which is subsequently simulated over \(t \in [0, N T]\), with \(N \sim 10^2\) in our current setup. 

The direct simulation for 50 timestep ($T/N_{data}$) in the current time unit will cost 400 GPU hours. With neural operator inference which takes only a few GPU sec, we achieve a speed up of $\sim10^5$x.

\textbf{Model Training.} 
With supervised manner, we train a neural operator (LocalNO~\cite{liu-schiaffini2024neural}) $\mathcal{F}_\theta$ to learn the mapping $u_t\to u_{t+\Delta T}$, where $u$ represents all the physical quantities in the system (appx. \ref{sec:method-simulation}). After training, we employ the neural operator and roll out autoregressively in the fine level, mimicking the quasi–steady state attained on small scales as $t \to \infty$.

The large-scale coupled dynamics studied in this work pose significant challenges for training:  
(1) extreme dynamic range in function values;  
(2) scarcity of training data;  
(3) singularities and highly non-uniform spatial distributions near the black hole center;  
(4) long-horizon rollouts in chaotic dynamics;  
(5) coupling across multiple physical fields and PDEs.

We introduce a suite of training techniques specifically designed to overcome these challenges.
The general principle is to incorporate physics prior to enhance the data efficiency and long-term stability.
(1) \textbf{Magnitude normalization}: we apply (signed-) logarithmic transform for quantities with large dynamic range, i.e., density, energy, and magnetic field. After that we apply robust z-scoring and soft clipping to normalize the data. (2) \textbf{Enforcing radial scaling}: It is known that energy and density display a radial scaling law: $\log u(\vec{x})\approx -k|\vec{x}|+b$, where $k$ can be estimated from the training data. In our experiments, the models are trained to predict the residual relative to this law: $\log(\texttt{Final Output})(\vec{x})=-k|\vec{x}|+\mathcal{N}^{-1}(\texttt{Model Output})(\vec{x})$, where $\mathcal{N}$ is the denormalization. This reduces the dynamic range the model needs to learn, improving accuracy and ensuring physical consistency.
(3) \textbf{Shell embedding}: 
Since the majority of the fluctuations of the functions occur near the black hole center, we employ a shell embedding to ensure the model to focus on the central region. We divide the domain into 8 regions based on the logarithmic distance to the center, and incoporate one-hot embedding of binary shell indicators $c\in\{0,1\}^8$.
We find this to perform better than standard position embedding.
(4) \textbf{Combined loss functions}: In addition to standard $L^2$ and $H^1$ losses for all eight quantities, we add regularizations on dissipation \cite{li2022learning} and the maximum of density and energy model prediction to prevent severe deviation from the radial scaling. A region-of-interest (ROI) mask further increases the loss weight near the black hole center~\cite{girshick2015fast}.

Training takes 10 GPU hours, far less than the cost of direct numerical simulations.
More technical details in this section and implementations are presented in \Cref{sec:method-simulation,sec:impl}.

\begin{figure}[t]
    \centering
    \includegraphics[width=0.95\linewidth]{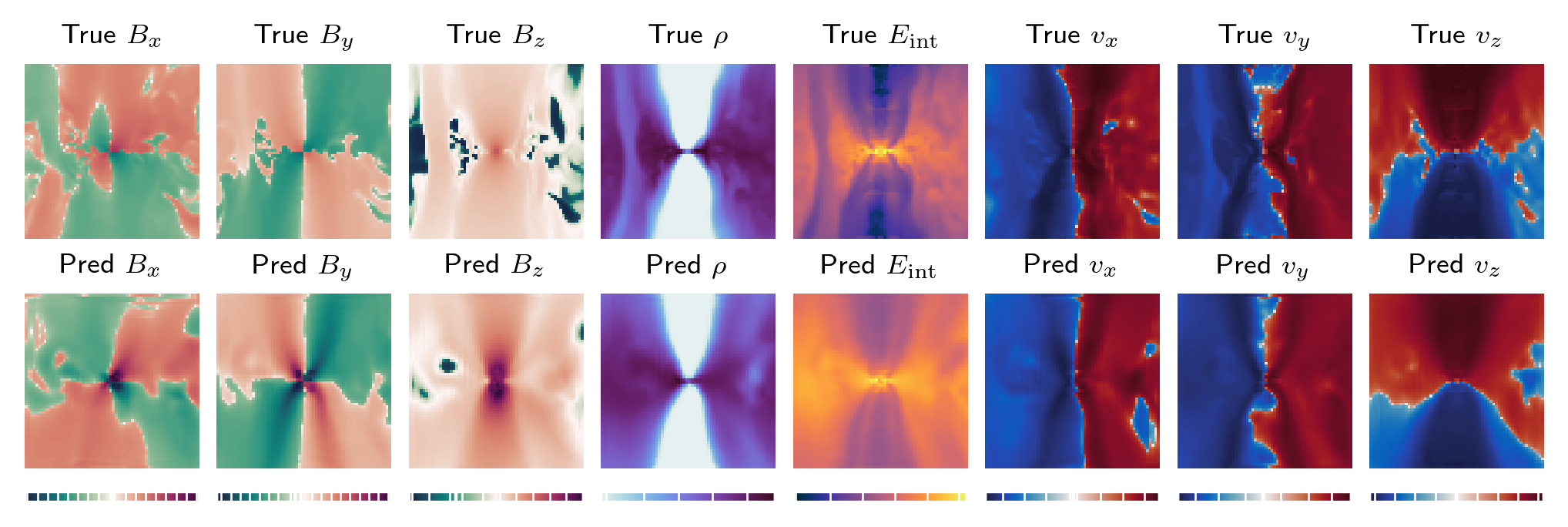}
    \caption{\textbf{GRMHD outputs.}
        Simulated and predicted magnetic field components \((B_x,B_y,B_z)\), density \(\rho\), internal energy \(E_{int}\), and velocity components \((v_x,v_y,v_z)\) from the GRMHD simulations.
        Jet structure near the polar region and disk structure near the midplane is preserved.
        }
    \label{fig:grmhd-outputs}
    \vspace{-1.8em}
\end{figure}

\vspace{-1.em}
\section{Results}
\vspace{-0.8em}
For both MHD and GRMHD, we train the model with the last 250 snapshots of the dataset and use an 80/20 train/validation split. We also train a CNN as baseline method, and conduct ablation study to validate the effect and necessity of training techniques we mentioned above. See appx. \ref{sec:ablation} for details.

\textbf{Evaluation.} Recall that the ultimate goal in this study is to develop a \textit{subgrid blackhole} providing realtime feedback for the large scale (coarse level), which requires long-term simulation in the fine level. Standard metric like $L^2$ relative error quantifying transient-state prediction does not make much sense here, as error will inevitably accumulate in these chaotic dynamics. In contrast, the most important merit is the ability to capture the fine-level statistical property well. As is the case, all the baseline learning methods in our ablation study and CNN achieve similar accuracy for per-step prediction, but none of these methods except the one we proposed generate unphysical artifacts after long rollouts (Table~\ref{tab:ablation-table}, Appx. \ref{sec:ablation}). Moreover, the absence of ground-truth (due to the intractable computation cost) for both long-term simulation in the fine level and simulation with real-time feedback in the coarse level suggests we can only make qualitative comparison and check certain observables to validate our method. 

\textbf{MHD}
Qualitative outputs are shown in Fig.~\ref{fig:mhd-outputs} (1). As boxed yellow in the figure, predictions made by CNN contain unphysical ripples in the magnetic field, and the velocity field does not match the torus structure. Our method derive physically faithful results. In Fig.~\ref{fig:mhd-outputs} (2), analytical prescribed subgrid scheme fails to capture the jet, demonstrating the importance of having a data-driven dynamical closure as in our approach. In Fig.~\ref{fig:mhd-outputs} (3), the observables from our method match the truth well. Note that in our sub-grid blackhole framework, the fine level supplies the boundary for coarse level, thus the model preidction is applicable as long as it provides accurate results near the outer boundary in the fine-level simulation.


\textbf{GRMHD}
Qualitative outputs are shown in Fig.~\ref{fig:grmhd-outputs}. Jet structure and central torus are preserved.

\textbf{Subgrid Black Hole in Simulation} The implementation detail and result for zooming back to the coarse-level are described in appx. \ref{apdx: insert back} and \cref{fig:ablation-vis}.

Our approach can serve as a first step (two-level version) towards a multi-level cyclic-zoom/multi-zone neural operator framework, enabling large-scale simulation with dynamic feeding-feedback that is previously intractable.

\textbf{Acknowledgement}
The authors are grateful for discussions with Philip Hopkins.
HYW thanks Minghao Guo for discussions on a closely related project.
ERM and HYW acknowledge support from the National Science Foundation through award NSF-AST2508940.
Part of the simulations were performed on DOE OLCF Summit under allocation AST198.
This research used resources of the Oak Ridge Leadership Computing Facility at the Oak Ridge National Laboratory, which is supported by the Office of Science of the U.S. Department of Energy under Contract No. DE-AC05-00OR22725. 
This work used Delta at the
National Center for Supercomputing Applications (NCSA) through allocation CIS250856 from the Advanced Cyberinfrastructure Coordination Ecosystem: Services \& Support (ACCESS) program~\cite{10.1145/3569951.3597559}, which is supported by U.S. National Science Foundation grants $\#$2138259, $\#$2138286, $\#$2138307, $\#$2137603, and $\#$2138296.

\bibliographystyle{unsrt}
\bibliography{cbd_ref,ml_ref}

\clearpage
\appendix
 \section*{Appendix}
The structure of the appendix is as follows.

\begin{itemize}
    \item \Cref{sec:method-simulation} describes dynamical systems of MHD and GRMHD, as well as their numerical simulations.
    \item \Cref{apdx:obsvb} summarizes the observables we considered in this work as an validation metric.
    \item \Cref{sec:impl} presents the detailed methods we adopt to train a neural operator under this large-scale astrophysics problem scenario, as well as the implementation details for model training.
    \item \Cref{sec:ablation} contains the experimental results for ablation studies and comparisons with baseline methods.
    \item \Cref{apdx: insert back} provides the detailed method for zoom-out part of our feeding-feedback paradigm and the results.
\end{itemize}

\section{Method-Simulation}
\label{sec:method-simulation}

We use a performance portable version of \texttt{Athena++}~\cite{Stone2020} based on \texttt{Kokkos} library~\cite{Trott2021}-\texttt{AthenaK}~\cite{athenak}. We use piecewise parabolic reconstruction \citep{Colella1984}, an HLLD Riemann solver \citep{Miyoshi2005},  a constraint transport algorithm \citep{Gardiner2008} for the divergence-free magnetic field evolution, and first-order flux correction \cite{2009ApJ...691.1092L}.

\subsection{Magneto-hydrodynamics Simulations: Magnetized Bondi Accretion}
The setup we used in this work is magnetized Bondi accretion onto a central supermassive black hole \cite{Guo2024}. Except for the case that the current run is without cooling and heating terms, the setup is the same as the fiducial run in \cite{Guo2024}. 
In this problem, we solve the Newtonian magnetohydrodynamics equations in conservative form:
\begin{align}
    \frac{\partial \rho}{\partial t}+\nabla \cdot(\rho \boldsymbol{v})                                                                                   & =s_{\rho}\,,                                \\
    \frac{\partial(\rho \boldsymbol{v})}{\partial t}+\nabla \cdot(\rho \boldsymbol{v} \boldsymbol{v} + P \boldsymbol{I}  - \boldsymbol{B}\boldsymbol{B}) & =\boldsymbol{s_p}-\rho \nabla \Phi\,,       \\
    \frac{\partial E}{\partial t}+\nabla \cdot[(E+{P}) \boldsymbol{v} - \left(\boldsymbol{B}\cdot \boldsymbol{v}\right) \boldsymbol{B}]                  & =s_E-\rho \boldsymbol{v} \cdot \nabla \Phi, \\
    \frac{\partial \boldsymbol{B}}{\partial t}-\nabla \times [ \boldsymbol{v} \times \boldsymbol{B} ]                                                    & =0\,,
\end{align}
where $\rho:\mathbb{R}^3\to\mathbb{R}$ is the gas density, $\boldsymbol{v}:\mathbb{R}^3\to\mathbb{R}^3$ is the velocity, $P:\mathbb{R}^3\to\mathbb{R}$ is the pressure, $E=P/(\gamma-1)+\rho |v|^2/2$ is the total energy density, and $-\nabla\Phi$ is the gravitational acceleration due to the central accretor. The flow evolved following ideal gas law with adiabatic index $\gamma=5/3$. The source terms $s_\rho$, $\boldsymbol{s_p}$, and $s_E$ represent removal of mass, momentum, and energy by the central black hole, respectively (e.g., accretion or feedback). These source terms are concentrated at the very center of the domain, covering $~1$ grid for the training set resolution.

In code unit, $GM=r_0=\rho_0=1$. The output quantities include $(\rho, P, \boldsymbol{v}=(v_x, v_y, v_z), \boldsymbol{B}=(B_x, B_y, B_z))$.

\subsection{GRMHD Simulations}
The conservative Valencia formulations for solving GRMHD equations with induction equation
\begin{align}
     & \partial_t\boldsymbol{U} + \partial_i\boldsymbol{F}=\boldsymbol{S} \\
     & \partial_t (\sqrt{-g}B^i)+\partial_j(\sqrt{-g}(b^iu^j-b^ju^i))=0
\end{align}
where the conserved variables, fluxes, and source terms associated with the connection are
\begin{equation}
        \mathbf{U} =\sqrt{-g}\left[\begin{array}{c}
                                             \rho u^t \\
                                             T_i^t    \\
                                             T_t^t+\rho u^t
                                         \end{array}\right],                                                  
        \mathbf{F} =\sqrt{-g}\left[\begin{array}{c}
                                             \rho u^j \\
                                             T_i^j    \\
                                             T_t^j+\rho u^j
                                         \end{array}\right],                                                  
        \mathbf{S} =\sqrt{-g}\left[\begin{array}{c}
                                             0                                                                    \\
                                             \frac{1}{2}\left(\partial_i g_{\alpha \beta}\right) T^{\alpha \beta} \\
                                             0
                                         \end{array}\right]
\end{equation}
respectively, where the metric is $\boldsymbol{g}$~($g=\det{\boldsymbol{g}}$). The rest-mass density is $\rho$, the coordinate frame 4-velocity is $u^\mu$ and the stress-energy tensor is
\begin{equation}
    {T^\mu}_{\nu} = w u^\mu u_\nu-b^\mu b_\nu + (p_g+p_m)\delta^\mu_\nu
\end{equation}
where the magnetic field is $b^\mu$, the magnetic pressure is $p_m= b_\mu b^\mu/2$, and the total enthalpy is $w$.
The non-monopole constraint is preserved when evolving the 3-components $B^i$ of the magnetic field:
\begin{equation}
    \partial_j (\sqrt{-g} B^j)=0
\end{equation}
where
\begin{equation}
    b^t = u_i B^i,                 
    b^i = \frac{1}{u^t}(B^i+b^tu^i).
\end{equation}
We solve the GRMHD equations in a Cartesian Kerr-Schild coordinate.
We adopt piecewise parabolic spatial reconstruction, HLLE Riemann solver, RK2 time integrator, and first-order flux correction \cite{2009ApJ...691.1092L}.

The setup we used in this work is the accretion of a Fishbone-Moncrief torus~\cite{1976ApJ...207..962F} onto a spinning black hole (with spin $a=0.9$). We choose the default magnetically arrested state setup in \texttt{AthenaK} public repo.

In all runs, we adopt $G=M=\rho_0=1$. The output quantities include $(\rho, P, \boldsymbol{v}=(v_x, v_y, v_z), \boldsymbol{B}=(B_x, B_y, B_z))$.

\section{Observables}\label{apdx:obsvb}
Here we use the spherically averaged mass density $\rho$, temperature $T=P/\rho$, and mass accretion rate $\dot M$ as observables (shown in panel (2) of Fig. \ref{fig:mhd-outputs}). The mass accretion rate is defined by 
\begin{equation}
    \dot{M} \equiv-\int_S \rho u^r\sqrt{-g}\,d \Omega,          
\end{equation}
In the Newtonian limit, $\sqrt{-g}=1$.

\section{Implementation Details}
\label{sec:impl}
    
\subsection{System and Environment}
Training is executed on Linux (x86\_64) with CPython~3.11 and CUDA.
A single NVIDIA GeForce RTX 4090 GPU is used; data loading employs pinned memory and worker parallelism.
Mixed precision is disabled by default; gradient accumulation is enabled.

\subsection{Data Representation}
Each snapshot contains eight 3D fields on a $64^3$ grid, stacked as a tensor of shape $(C,D,H,W)$ with $C{=}8$: magnetic–field components $B_x,B_y,B_z$ (variables \texttt{bcc1}, \texttt{bcc2}, \texttt{bcc3}), mass density $\rho$ (\texttt{dens}), internal energy $e$ (\texttt{eint}), and velocity components $v_x,v_y,v_z$ (\texttt{velx}, \texttt{vely}, \texttt{velz}).

\subsection{Normalization: Channel-wise Transforms, Robust Scaling, and Soft Clipping}
\paragraph{Transforms.}
Before standardization, each channel $x_c$ is mapped by a type-specific $T_c$:
\[
\hat{x}_c \!=\!
\begin{cases}
\log_{10}\!\bigl(x_c+\varepsilon_c\bigr), & \texttt{pos} \;\;(\texttt{dens},\texttt{eint}),\\[3pt]
\sgn(x_c)\,\log_{10}\!\Bigl(1+\tfrac{|x_c|}{\varepsilon_c}\Bigr), & \texttt{signed} \;\;(\texttt{bcc1:3}),\\[6pt]
x_c, & \texttt{linear} \;\;(\texttt{velx:velz}).
\end{cases}
\]
Per-channel scales $\varepsilon_c$ are estimated from a representative training subset:
\[
\varepsilon_c =
\begin{cases}
10^{\lfloor \log_{10}(\min x_c)\rfloor-2}, & \texttt{pos},\\[3pt]
10^{\lfloor \log_{10}(\max |x_c|)\rfloor-2}, & \texttt{signed},\\[3pt]
0, & \texttt{linear}.
\end{cases}
\]

\paragraph{Robust z-scoring and soft clipping.}
Let $m_c=\mathrm{median}(\hat{x}_c)$ and $s_c = 1.4826\,\mathrm{median}\!\bigl(|\hat{x}_c-m_c|\bigr)$ (clamped to $10^{-6}$).
Define
\[
z_c=\frac{\hat{x}_c-m_c}{s_c},
\qquad
\tilde{z}_c = \gamma\,\tanh\!\Bigl(\frac{z_c}{\gamma}\Bigr),
\]
with soft-clip range $\gamma{=}6$ (default).
Decoding inverts the pipeline via
\[
z_c = \gamma\,\mathrm{atanh}\!\Bigl(\frac{\tilde{z}_c}{\gamma}\Bigr),\quad
\hat{x}_c = s_c z_c + m_c,\quad
x_c = T_c^{-1}(\hat{x}_c),
\]
where
\[
T_c^{-1}(u)=
\begin{cases}
10^{u}-\varepsilon_c, & \texttt{pos},\\[3pt]
\sgn(u)\,\varepsilon_c\!\left(10^{|u|}-1\right), & \texttt{signed},\\[3pt]
u, & \texttt{linear}.
\end{cases}
\]
Because of soft clipping, the inverse is approximate near $|\tilde{z}_c|\approx \gamma$; we clamp inputs to $\mathrm{atanh}$ to $0.99\,\gamma$.

\subsection{Radial Shell Positional Channels}
To provide absolute radial context, we augment inputs with $m$ concentric \emph{shell} indicators on the $64^3$ grid.
Let the grid center be $c$ and define $r(i,j,k)=\| (i,j,k)-c\|_2$.
Choose logarithmically spaced boundaries $0=b_0 < b_1 < \cdots < b_m = r_{\max}$ on $[10^{-1},r_{\max}]$ (defaults: $m{=}8$, $r_{\max}{=}10$).
Each voxel receives an index $s(i,j,k)$ with $b_{s}\le r < b_{s+1}$ (voxels with $r\ge r_{\max}$ use $s{=}m{-}1$).
We one-hot encode into $E\in\{0,1\}^{m\times 64\times 64\times 64}$ and \emph{concatenate} $E$ to the eight physical channels.
Shell channels are binary and are never normalized; normalization applies only to the physical channels.

\subsection{Single-Stage Radial Scaling Baseline (Residualization)}
We add a fixed radial reference to the \texttt{dens} and \texttt{eint} outputs and train the model to predict the residual.
With a precomputed radius grid $r$ on $64^3$, the reference is
\[
B(r) = k\,r,
\]
where the slope $k$ is estimated offline by least-squares regression of $\log_{10}(U)$ versus $r$ using $U=\texttt{dens}+\texttt{eint}$.
During inference, the wrapped predictor returns $\hat{y}=f_\theta(x)+B(r)$ on the \texttt{dens}/\texttt{eint} channels.

\paragraph{Residual envelope around the radial baseline.}
With the reference \(B(r)\) applied to \texttt{dens}/\texttt{eint}, the network predicts residuals \(\eta_c=\hat{y}_c - B(r)\) for \(c\in\{\rho,e\}\).
Since \(B(r)\) captures the global radial decay, \(\eta_c\) should be near–zero and encode only local, non-radial structure.
To prevent the model from re-learning the radial trend or drifting by large offsets, we impose a symmetric hinge “envelope” of half-width \(\Delta_c\) (in normalized units) around \(B(r)\):
\[
\mathcal{L}_{\mathrm{env}}
= w_\rho\,\mathbb{E}\!\bigl[\bigl(\,|\hat{y}_{\rho}-B(r)|-\Delta_\rho\,\bigr)_+^{\!2}\bigr]
+ w_e\,\mathbb{E}\!\bigl[\bigl(\,|\hat{y}_{e}-B(r)|-\Delta_e\,\bigr)_+^{\!2}\bigr],
\]
where \((u)_+=\max(u,0)\).
The envelope is radius-aware (centered at \(B(r)\)), gives zero penalty to residuals within \(\pm\Delta_c\) so fine structure can be learned, and grows quadratically outside to discourage large bias.
We choose \(\Delta_c\) in normalized coordinates (fixed hyperparameters; e.g., \(\Delta_\rho=\Delta_e=1.5\) unless noted), roughly matching the robust spread of residuals on the training set.

\subsection{Data-Driven Physical Bounds and Mapping to Normalized Space}
We compute robust bounds for density and internal energy on the \emph{training partition} using quantiles $q_{\rm low},q_{\rm high}$ (defaults: $0.001,0.999$).
With a symmetric multiplicative margin $\delta$ (default: $5\%$) in physical units,
\[
\ell_\rho=(1-\delta)\,Q_{q_{\rm low}}(\rho),\quad
u_\rho=(1+\delta)\,Q_{q_{\rm high}}(\rho),\qquad
\ell_e=(1-\delta)\,Q_{q_{\rm low}}(e),\quad
u_e=(1+\delta)\,Q_{q_{\rm high}}(e).
\]
Bounds are mapped to the normalized domain via each channel’s transform and robust scaling:
\[
b^{\rm norm}=\frac{T_c(b)-m_c}{s_c}.
\]
During training we penalize violations wrt.\ these normalized thresholds; at evaluation we clamp \texttt{dens}/\texttt{eint} in normalized space to $[\ell^{\rm norm},u^{\rm norm}]$ prior to decoding.

\subsection{Objective Function}
Let $\hat{y}$ and $y$ denote predictions and targets in the normalized domain, with channels grouped as magnetic field $\mathbf{B}{:}{=}\texttt{bcc1:3}$, velocity $\mathbf{v}{:}{=}\texttt{velx:velz}$, density $\rho$, and internal energy $e$.

\paragraph{Component-weighted $L^2$.}
\[
\mathcal{L}_{\mathbf{B}}=\lambda_B\|\hat{y}_{\mathbf{B}}-y_{\mathbf{B}}\|_2^2,\quad
\mathcal{L}_{\mathbf{v}}=\lambda_{\rm vel}\|\hat{y}_{\mathbf{v}}-y_{\mathbf{v}}\|_2^2,\quad
\mathcal{L}_{\rho}=\|\hat{y}_{\rho}-y_{\rho}\|_2^2,\quad
\mathcal{L}_{e}=\|\hat{y}_{e}-y_{e}\|_2^2.
\]

\paragraph{$H^1$ match.}
\[
\mathcal{L}_{H^1}=\lambda_{H^1}\sum_{c}\|\nabla \hat{y}_c-\nabla y_c\|_2^2.
\]

\paragraph{Velocity ROI emphasis.}
Let $v=\mathrm{decode}(y_{\mathbf{v}})$ and $s=\|v\|_2$ (voxelwise speed).
For each batch, define a threshold $\tau$ as the $(100-\rho)\%$ quantile of $s$ (default $\rho{=}20$).
With mask $M=\mathbf{1}[s\ge\tau]$, scale $\kappa{=}8$, and linear epoch ramp $r(e)=\min(1,e/375)$,
\[
\mathcal{L}_{\rm vel,ROI}
= r(e)\,\kappa\ \mathbb{E}\!\left[\frac{\|(\hat{y}_{\mathbf{v}}-y_{\mathbf{v}})\odot M\|_2}{\|y_{\mathbf{v}}\odot M\|_2+\epsilon}\right].
\]

\paragraph{Dissipative regularizer.}
To discourage spurious amplification, we penalize \emph{only} positive growth of the global $L^2$ norm in normalized space. Let $\|\cdot\|$ denote the $L^2$ norm over all channels/voxels. From the training set compute $R_{\max}=\max\|x\|$ and set
$R_{\rm in}=1.05\,R_{\max}$ and $R_{\rm out}=1.5\,R_{\rm in}$. Define a contracted target
$y_{\rm tgt}=(R_{\rm in}/R_{\rm out})\,x$.
For an input $x$ with raw prediction $y_{\rm pred}$, use a smooth gate
$\rho(x)=\sigma\!\big(\beta\,[R_{\rm in}-\|x\|]\big)$ (with $\beta{=}10$) and blend
$\tilde y=\rho(x)\,y_{\rm pred}+\big(1-\rho(x)\big)\,y_{\rm tgt}$.
The loss is
\[
\mathcal{L}_{\rm diss}=\alpha\,\mathbb{E}\!\left[\big(\,\|\tilde y\|-\|x\|\,\big)_{+}\right],
\qquad \alpha=5{\times}10^{-4},\ \ (u)_{+}=\max(u,0).
\]
Intuitively, states with $\|x\|\!\ll\!R_{\rm in}$ are unaffected ($\rho\!\approx\!1$), while large-amplitude states are softly nudged towards a uniformly contracted copy.

\paragraph{Constraint penalties (normalized domain).}
\begin{align}
\mathcal{L}_{\rm constr} &=
\lambda_e^{\rm lo}\,\mathbb{E}\!\bigl[\max\{0,\ \ell_e^{\rm norm}-\hat{y}_e\}^2\bigr]
+ \lambda_\rho^{\rm lo}\,\mathbb{E}\!\bigl[\max\{0,\ \ell_\rho^{\rm norm}-\hat{y}_\rho\}^2\bigr]\\
&+ \lambda_\rho^{\rm hi}\,\mathbb{E}\!\bigl[\max\{0,\ \hat{y}_\rho-u_\rho^{\rm norm}\}^2\bigr]
+ \lambda_e^{\rm hi}\,\mathbb{E}\!\bigl[\max\{0,\ \hat{y}_e-u_e^{\rm norm}\}^2\bigr].    
\end{align}

\paragraph{Total loss.}
\[
\mathcal{L}=
\underbrace{\mathcal{L}_{\mathbf{B}}+\mathcal{L}_{\mathbf{v}}+\mathcal{L}_{\rho}+\mathcal{L}_{e}}_{\text{component-weighted }L^2}
+\lambda_{H^1}\mathcal{L}_{H^1}
+\mathcal{L}_{\rm vel,ROI}
+\mathcal{L}_{\rm dissip}
+\mathcal{L}_{\rm env}
+\mathcal{L}_{\rm constr}.
\]

\subsection{Backbone and Wrapper}
\paragraph{Backbone.}
We employ a 3D \emph{Local Neural Operator} with equidistant discrete--continuous convolutions (DISCO) \cite{liu-schiaffini2024neural} specialized to volumetric inputs.
When shell labels are enabled, inputs have $8{+}m$ channels, where $m$ comes from shell embedding; outputs remain the eight physical channels.

\paragraph{Wrapper.}
A \emph{Scaling Baseline Wrapper} adds $B(r)=k\,r$ to the \texttt{dens}/\texttt{eint} channels on top of the base model’s residual prediction.

\subsection{Optimization and Schedule}
\paragraph{Defaults.}
\begin{itemize}[leftmargin=*]
\item \textbf{Epochs:} 1200;\quad \textbf{Batch size:} 4;\quad \textbf{Gradient accumulation:} 4 (effective batch 16).
\item \textbf{Optimizer:} Adam, learning rate $10^{-3}$, weight decay $10^{-4}$.
\item \textbf{LR schedule:} Linear warmup for 75 epochs from $0.1\times$lr, then cosine annealing to $10^{-6}$.
\item \textbf{Gradient clip:} global norm $1.0$;\quad \textbf{Early stopping:} patience 100 epochs.
\item \textbf{Normalization clip range:} $\gamma{=}6.0$.
\item \textbf{Shell labels:} enabled; $m{=}8$, $r_{\max}{=}10$.
\item \textbf{Scaling baseline:} enabled; slope $k=-1.78\times 10^{-2}$ (single-stage).
\item \textbf{Quantile bounds:} $q_{\rm low}{=}0.001$, $q_{\rm high}{=}0.999$, margin $\delta{=}5\%$.
\item \textbf{Velocity ROI:} top $20\%$, scale $8.0$, ramp $375$ epochs.
\item \textbf{Loss weights:} $\lambda_B{=}1.2$, $\lambda_{\rm vel}{=}1.0$, $\lambda_{H^1}{=}0.05$, $\alpha_{\rm dissip}{=}5\times 10^{-4}$, residual envelope $w_\rho{=}w_e{=}0.05$ with $\Delta_\rho{=}\Delta_e{=}1.5$, constraint floors $\lambda_e^{\rm lo}{=}\lambda_\rho^{\rm lo}{=}0.05$, upper-bound weights off by default.
\end{itemize}

\begin{table}[t]
\centering
\caption{Default training configuration (summary).}
\vspace{3pt}
\begin{tabular}{l l}
\toprule
\textbf{Parameter} & \textbf{Value} \\
\midrule
Epochs & 1200 \\
Batch size / Grad.\ accumulation & $4$ / $4$ (effective $16$) \\
Adam learning rate / Weight decay & $1.0\times 10^{-3}$ / $1.0\times 10^{-4}$ \\
LR schedule & Warmup $75$ $\rightarrow$ Cosine to $10^{-6}$ \\
Gradient clip (global norm) & $1.0$ \\
Early stopping patience & $100$ \\
Normalization clip range $\gamma$ & $6.0$ \\
Shell labels & enabled, $m{=}8$, $r_{\max}{=}10$ \\
Scaling baseline & single-stage, $k=-1.78\times 10^{-2}$ \\
Quantile bounds $(q_{\rm low},q_{\rm high})$ & $(0.001,\,0.999)$, margin $5\%$ \\
Velocity ROI & top $20\%$, scale $8$, ramp $375$ \\
$L^2$ weights & $\lambda_B{=}1.2$, $\lambda_{\rm vel}{=}1.0$ \\
$H^1$ & $\lambda_{H^1}{=}0.05$ \\
Dissipative regularizer & $\alpha_{\rm dissip}{=}5\times 10^{-4}$ \\
Residual envelope & $\Delta_\rho{=}\Delta_e{=}1.5$, $w_\rho{=}w_e{=}0.05$ \\
Constraint penalties & $\lambda_e^{\rm lo}{=}\lambda_\rho^{\rm lo}{=}0.05$; upper off \\
\bottomrule
\end{tabular}
\label{tab:config}
\end{table}

\subsection{Metrics, Logging, and Evaluation}
\paragraph{Validation metric.}
Unless noted otherwise, we report volumetric relative $L_2$ error (\%) on the validation set in the normalized domain (after transforms and soft clipping), either per-channel or average across channels.

\paragraph{Diagnostics and logging.}
Per-batch diagnostics include tensor shapes and dtype, global min/max/mean/std, NaN/Inf counts, gradient norms at regular intervals, and rates of constraint violations.
Loss components and scheduler state are logged throughout training.

\paragraph{Evaluation-time clamping and decoding.}
For \texttt{dens}/\texttt{eint}, predictions are clamped in normalized space to $[\ell^{\rm norm},u^{\rm norm}]$ before decoding to physical units.
We then compute spectra and multi-step rollout diagnostics as applicable.

\subsection{Practical Notes}
\begin{itemize}[leftmargin=*]
\item \textbf{Soft-clip inversion is approximate.} Values near the clip boundary do not round-trip exactly; inputs to $\mathrm{atanh}$ are clamped to $0.99\,\gamma$.
\item \textbf{Bounds depend on the training partition.} Quantile bounds are computed on training data to prevent leakage; changing the split or $\delta$ will move thresholds.
\item \textbf{Normalized-space semantics.} Envelope widths and penalties are specified in z-scored units; their physical meaning depends on $m_c$ and $s_c$.
\item \textbf{Channel dimensionality with shells.} Enabling shell labels increases input channels from $8$ to $8{+}m$; outputs remain the eight physical channels.
\end{itemize}


\section{Ablation Study}
\label{sec:ablation}

\paragraph{Protocol (held constant).}
Unless stated otherwise, all settings match Sec.~\ref{sec:impl}:
same data split, canonical $64^3$ grid, channel--wise transforms with robust z-scoring and soft clipping, optimization/schedule (epochs, batch, LR, warmup+cosine, gradient clipping, early stopping), and reporting in the normalized domain.
Decoding and evaluation-time clamping for \texttt{dens}/\texttt{eint} follow the same procedure.

\subsection{Ablation configurations}
We ablate one component at a time, keeping everything else as in the default setup (“\emph{Ours}”).
Let $\mathbf{B}{:=}\texttt{bcc1:3}$, $\mathbf{v}{:=}\texttt{velx:velz}$, $\rho{:=}\texttt{dens}$, $e{:=}\texttt{eint}$, and let $B(r){=}k\,r$ denote the single-stage radial baseline applied on the $\rho/e$ channels.

\begin{itemize}[leftmargin=*]
\item \textbf{Plain L2.}
Remove all auxiliary loss terms and weights; minimize an \emph{unweighted} per-voxel MSE across the eight channels in the normalized domain:
\[
\mathcal{L}_{\text{plain}} \;=\; \sum_{c\in\{\mathbf{B},\mathbf{v},\rho,e\}} \|\hat{y}_c - y_c\|_2^2.
\]
Concretely, set $\lambda_B{=}\lambda_{\rm vel}{=}1$, disable $\mathcal{L}_{H^1}$, ROI velocity, dissipative regularizer.
\item \textbf{No radial/constraint.}
Keep component-weighted $L^2$, $H^1$, ROI velocity, and the dissipative regularizer as in Sec.~\ref{sec:impl}, but \emph{disable any radius-specific priors/constraints}:
(i) remove the scaling baseline $B(r)$ (train/predict in the raw normalized space),
(ii) remove residual envelope penalties on $\rho/e$,
(iii) remove constraint penalties (and do not clamp at evaluation).
Positional channels remain as in the default.
\item \textbf{No PE/Radial Shell.}
Remove all positional encodings: do not concatenate shell one-hots and do not use any alternative positional features.
Also set the inner-region loss nudge to zero so there is no implicit radial reweighting; keep all other terms (component-weighted $L^2$, $H^1$, ROI velocity, dissipative regularizer, constraints, and the single-stage $B(r)$).
\item \textbf{PE (Fourier positional encoding).}
Replace shell one-hots with Fourier features of \emph{normalized Cartesian coordinates} $(\xi,\eta,\zeta)\in[0,1]^3$:
for $\ell=0,\dots,L{-}1$ with $L{=}4$,
\[
\phi_\ell(u)=\bigl[\sin(2\pi 2^\ell u),\;\cos(2\pi 2^\ell u)\bigr],
\quad
u\in\{\xi,\eta,\zeta\}.
\]
Concatenate the $3\times 2L$ channels to the physical inputs; do not use shell one-hots.
All losses, constraints, and $B(r)$ remain enabled.
\item \textbf{CNN backbone (3D CNN instead of neural operator).}
We replace the Local Neural Operator with a lightweight 3D U\mbox{-}Net–style encoder–decoder matched for parameter count (base width 32; four resolution levels with channel widths $[32,64,128,256]$). Each stage uses residual $3{\times}3{\times}3$ conv blocks with GroupNorm and SiLU activations; downsampling is $\times 2$ via max-pooling and upsampling via transposed convolutions. Inputs are the eight physics channels $+\,n_{\text{shell}}$ radial shell one-hots; outputs are the physics channels. All training losses, constraints, and the radial baseline $B(r)$ are kept identical to the operator models.
\end{itemize}

\subsection{What each ablation isolates}
\begin{itemize}[leftmargin=*]
\item \emph{Plain L2} probes the net contribution of all auxiliary inductive biases beyond plain fidelity.
\item \emph{No radial/constraint} isolates the effect of the scaling prior and data-driven bounds from the rest of the loss design.
\item \emph{No PE/Radial Shell} measures the value of absolute positional information.
\item \emph{PE} compares Fourier PE against our shell-based positional encoding under otherwise identical training.
\item \emph{CNN backbone} tests whether the gains are architectural (operator layer) or purely due to losses/normalization.
\end{itemize}

\begin{figure}[t]
  \centering
  \includegraphics[width=\linewidth,trim=10mm 10mm 10mm 10mm,clip]{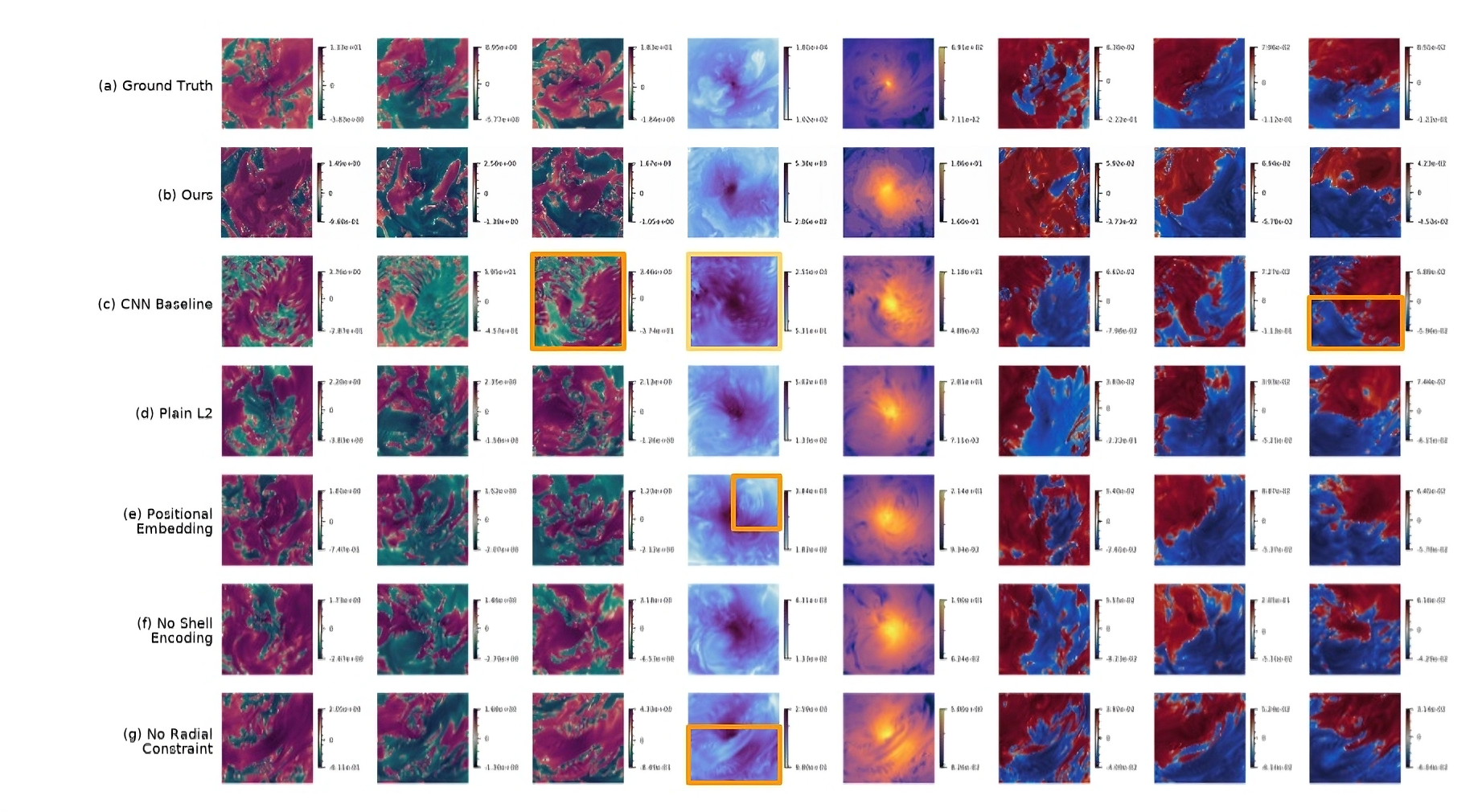}
  \caption{\textbf{Ablation study on a central slice at step $t{=}50$.}
Columns (left$\to$right): $B_x$, $B_y$, $B_z$, mass density $\rho$, internal energy $e$, and velocity components $(v_x,v_y,v_z)$. 
Values are in the model’s normalized domain (see Methods).
Rows: (a) \emph{Ground truth}; (b) \emph{Ours} (LocalNO + shell positional encoding + radial baseline/constraints + ROI and $H^1$ terms); (c) \emph{CNN baseline} (pure convolutional surrogate); (d) \emph{Plain $L^2$} (no component weights/ROI/$H^1$); (e) \emph{Positional Embedding only} (Fourier features, no shells); (f) \emph{No Shell Encoding}; (g) \emph{No Radial Constraint/Baseline}.
Colored boxes mark typical failure modes observed in ablations: \textbf{yellow}: over-smoothing/texture blur; \textbf{orange}: spurious mass/energy (unphysical deficits or excess). 
The full model (b) most closely matches the ground truth, preserving small-scale magnetic structure and inner-torus morphology while avoiding excessive smoothing and mass artifacts.}
  \label{fig:ablation-vis-50}
\end{figure}

\subsection{Rollout for 50 $\Delta T$ in MHD (\cref{fig:ablation-vis-50})}

\begin{figure}[t]
  \centering
  \includegraphics[width=\linewidth,trim=10mm 10mm 10mm 10mm,clip]{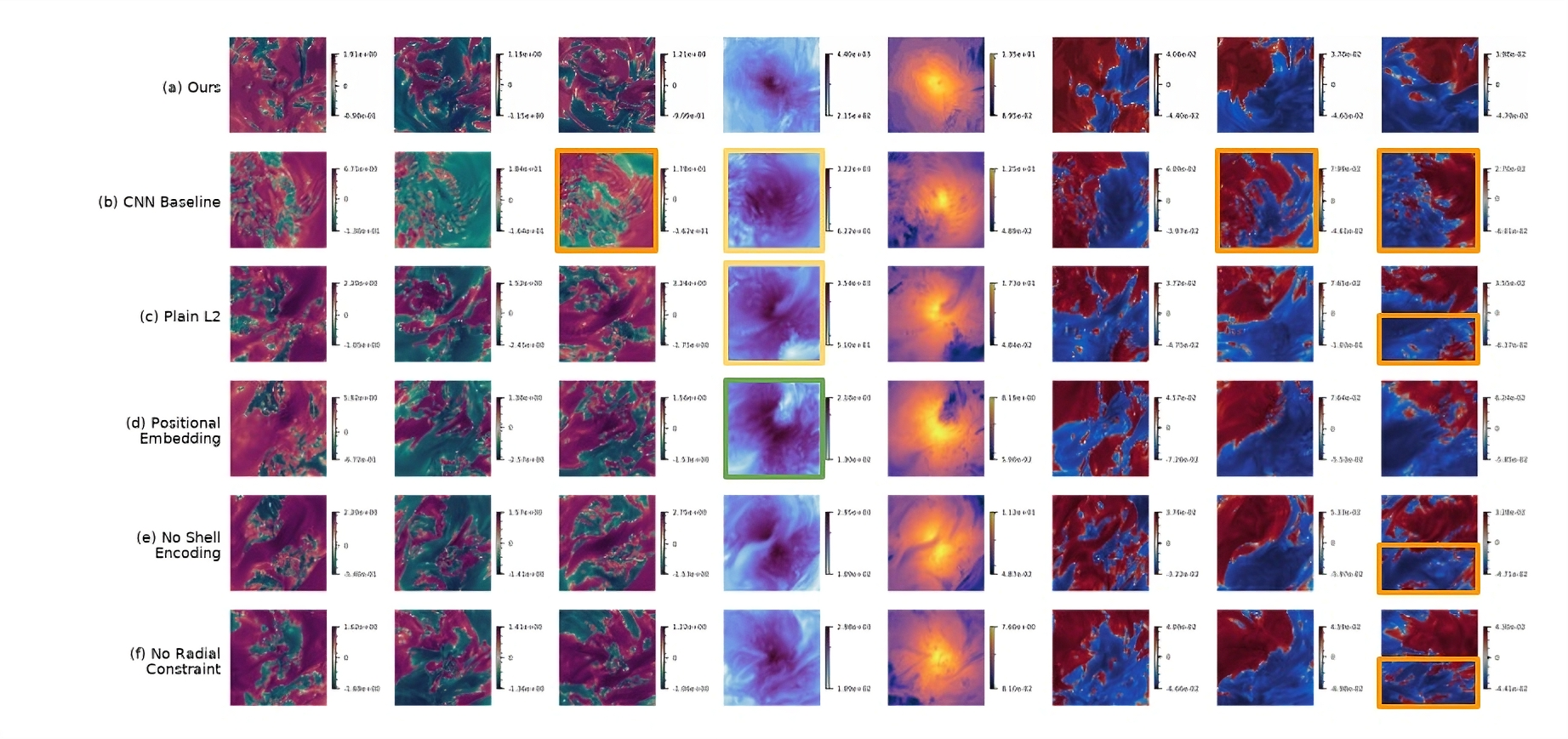}
  \caption{\textbf{Ablation study on a central slice at step $t{=}100$.}
Columns (left$\to$right): $B_x$, $B_y$, $B_z$, mass density $\rho$, internal energy $e$, and velocity components $(v_x,v_y,v_z)$. 
Values are in the model’s normalized domain (see Methods).
Rows: (a) \emph{Ours} (LocalNO + shell positional encoding + radial baseline/constraints + ROI and $H^1$ terms); (b) \emph{CNN baseline} (pure convolutional surrogate); (c) \emph{Plain $L^2$} (no component weights/ROI/$H^1$); (d) \emph{Positional Embedding only} (Fourier features, no shells); (e) \emph{No Shell Encoding}; (f) \emph{No Radial Constraint/Baseline}.
Colored boxes mark typical failure modes observed in ablations: \textbf{yellow}: over-smoothing/texture blur; \textbf{green}: anisotropic diffusion of the torus; \textbf{orange}: spurious mass/energy (unphysical deficits or excess). 
The full model (a) most closely maintains proper structure, preserving small-scale magnetic structure and inner-torus morphology while avoiding excessive smoothing and mass artifacts.}
  \label{fig:ablation-vis-100}
\end{figure}

\subsection{Rollout for 100 $\Delta T$ in MHD (\cref{fig:ablation-vis-100})}

\begin{table}[t]
\centering
\caption{This table reports the avg. relative $L_2$ error (\%) across all channels, along with the per-channel relative $L_2$ error (\%) based on the validation set.}
\label{tab:ablation-table}

\begingroup
\setlength{\tabcolsep}{3.5pt} 
\renewcommand{\arraystretch}{0.95} 
\begin{adjustbox}{max width=\linewidth}
\begin{tabular}{@{}lccccccccc@{}} 
\toprule
\textbf{Configuration} & \textbf{Avg. Error (\%)} &
$B_x$ (\%) & $B_y$ (\%) & $B_z$ (\%) & $\rho$ (\%) & $e$ (\%) & $v_x$ (\%) & $v_y$ (\%) & $v_z$ (\%)\\
\midrule
Ours                & 14.02 & 16.71 & 17.01 & 14.73 & 10.98 & 11.25 & 13.47 & 14.04 & 13.94 \\
PE (Fourier)        & 13.87 & 16.77 & 17.06 & 14.87 &  9.85 & 10.42 & 13.73 & 14.08 & 14.15 \\
No PE/Radial Shell  & 13.93 & 16.84 & 17.15 & 14.89 & 10.03 & 10.42 & 13.87 & 14.18 & 14.06 \\
No radial/constraint& 14.17 & 16.84 & 17.26 & 14.87 & 10.87 & 11.33 & 13.82 & 14.29 & 14.07 \\
Plain L2            & 13.69 & 16.76 & 16.98 & 14.67 &  9.55 &  9.89 & 13.45 & 14.06 & 14.16 \\
CNN backbone        & 19.09 & 23.85 & 23.91 & 21.44 & 15.21 & 14.87 & 17.46 & 17.80 & 18.17 \\
\bottomrule
\end{tabular}
\end{adjustbox}
\endgroup
\end{table}

\subsection{$L^2$ Loss while preserving dynamics near the singularity}
When omitting certain components in our ablation study, we notice a lower L2 validation error for density and internal energy, yet observe poor adherence to the ground truth for all fields at or directly around the black hole. The gravitational effect of the singularity creates a localized region of increased complexity, marginal in size when compared to the entire domain, causing it to be neglected by the Neural Operator in the objective to minimize overall loss. In such a case, we risk losing dynamics near the inner torus region and physics critical to the entire system, emphasizing the need for radial awareness and weightage.

\subsection{Reporting details}
For each row in Table~\ref{tab:ablation-table}, train with the same optimizer and schedule, and evaluate on the same validation split.
Unless explicitly disabled by the ablation, keep: component-weighted $L^2$, $H^1$ loss, ROI velocity emphasis, dissipative regularizer, data-driven bounds with evaluation-time clamping, robust normalization, soft clipping, and the single-stage baseline $B(r){=}k r$.

\begin{figure}[t]
  \centering
  \includegraphics[width=0.65\linewidth]{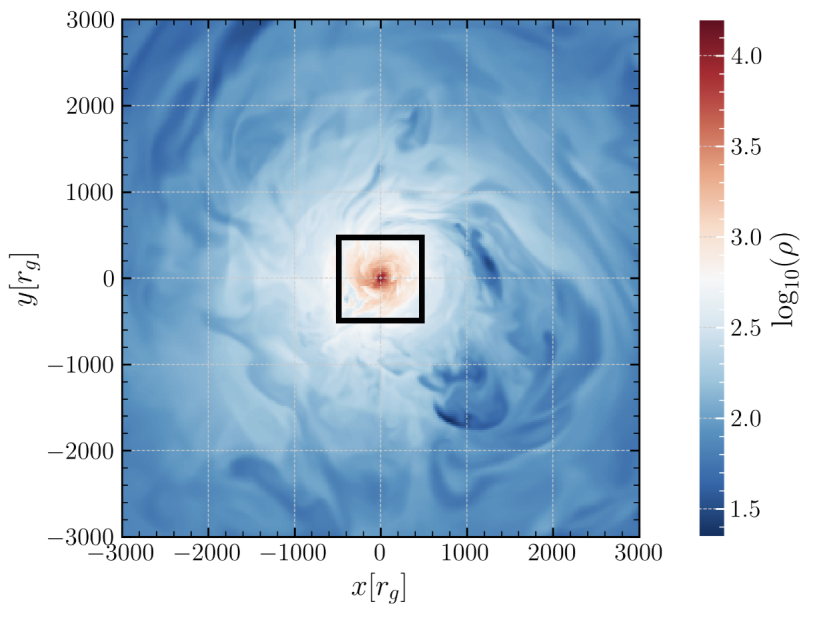}
  \caption{
Two-dimensional slice of mass density after coupling the neural operator back to the code. 
The subgrid region (roll-out of the neural operator) is denoted by the black square.
}
\label{fig:ablation-vis}
\end{figure}

\begin{figure}
    \centering
    \includegraphics[width=0.3\linewidth]{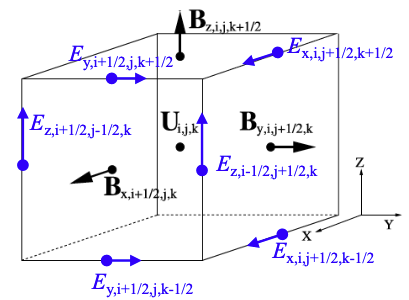}
    \caption{
    Positions of hydrodynamic variables, magnetic fields, and electric fields. The hydrodynamic variables are at the body-center of each cell, while the electric fields and magnetic fields are at the face-centers and edge-centers.
    } 
    \label{fig:ct}
\end{figure}

\begin{figure}
    \centering
    \includegraphics[width=0.4\linewidth]{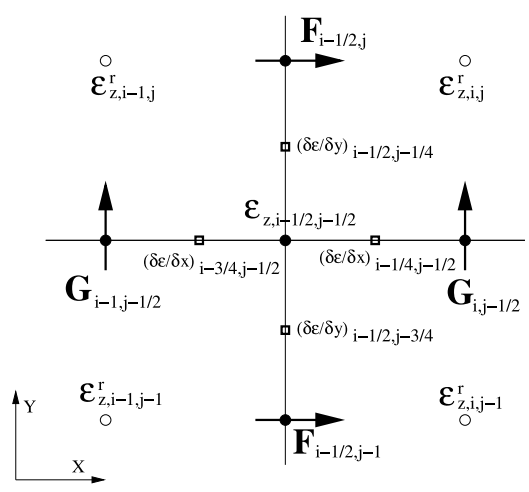}
    \caption{
    The schematic diagram of the x-y slice, in which shows the cell-centered reference state of EMF $E^r_z$, cell-cornered EMF $E_z$ used for recovering the area-averaged magnetic field, and the fluxes of conserved variables in the x and y directions.
    } 
    \label{fig:ct-2}
\end{figure}

\section{Coupling NO to Simulation}\label{apdx: insert back}
We briefly describe how we couple the neural operator to the direct simulation at coarser level in this section.
First, we directly output the rollout from neural operator as hdf5 files and each time read in two files to interpolate between two timesteps linearly.
The hydrodynamic variables $(\rho, P, v_x, v_y, v_z)$ are used to overwrite the corresponding values at the inner boundary directly. 
The treatment of coupling the magnetic field to the inner boundary is discussed in detail below.

\subsection{Review: Divergence-Free Magnetic Field and Constraint Transport}
The magnetic fields are evolved via Stoke's law:
\begin{equation}
\frac{\partial}{\partial t} \int_S \boldsymbol{B} \cdot d \boldsymbol{S}=-\int_L \boldsymbol{E} \cdot d \boldsymbol{l} \,,
\end{equation}
in the differential form : 
\begin{equation}
\frac{\partial \boldsymbol{B}}{\partial t}+\nabla \times \boldsymbol{E} =0 \,,
\end{equation}
where the electric field (electromotive force [EMF]) $\boldsymbol{E}=-\boldsymbol{v}\times \boldsymbol{B}$ in ideal MHD.

Constraint transport (CT) algorithm update the area-averaged magnetic fields using the line-averaged EMF at cell corners. First, we can separately define the face-centered, area-averaged magnetic field $\left(B_x\right)_{i+1 / 2, j, k}$ and the edge-centered, line-averaged EMF $\left(E_x\right)_{i, j+1 / 2, k-1 / 2}$ as follow (taking $x$-component as an example): 
\begin{align}
&\left(B_x\right)_{i+1 / 2, j, k} =\frac{1}{\Delta y \Delta z} \int_S B_x(y, z) d y d z\, , \\
&\left(E_x\right)_{i, j+1 / 2, k-1 / 2} =\frac{1}{\Delta x \Delta t} \int E_x(x) d x d t\, ,
\end{align}

The magnetic field at the $n+1$ timestep is updated using the electro-magnetic forces accordingly
\begin{align}
B_{x, i+1 / 2, j, k}^{n+1}=B_{x, i+1 / 2, j, k}^n-&\frac{\Delta t}{\Delta y}\left(E_{z, i-1 / 2, j+1 / 2, k}^{n+1 / 2}-E_{z, i-1 / 2, j-1 / 2, k}^{n+1 / 2}\right)\\
+&\frac{\Delta t}{\Delta z}\left(E_{y, i-1 / 2, j, k+1 / 2}^{n+1 / 2}-E_{z, i-1 / 2, j, k-1 / 2}^{n+1 / 2}\right)
\end{align}
Then the divergence of magnetic field $\nabla\cdot \boldsymbol{B}$ is preserved to machine accuracy
\begin{equation}
\nabla \cdot \boldsymbol{B}  =\frac{B_{x, i+1 / 2, j, k}-B_{x, i-1 / 2, j, k}}{\Delta x} 
 +\frac{B_{y, i, j+1 / 2, k}-B_{x, i, j-1 / 2, k}}{\Delta y} 
 +\frac{B_{z, i, j, k+1 / 2}-B_{z, i, j, k-1 / 2}}{\Delta z}
\end{equation}

\subsection{Coupling NO with Constraint Transport}
In \texttt{Athena++/K}, it is the Riemann solver that returns area-averaged electric fields at cell faces, while the CT requires the line-averaged EMF at cell corners to update the area-averaged magnetic fields, here this is achieved by \cite{Gardiner:2005JCoPh.205..509G,Gardiner2008,Stone:2008mh}:
\begin{align}
E_{z, i-1 / 2, j-1 / 2}= 
&\frac{1}{4}\left(E_{z, i-1 / 2, j}+E_{z, i-1 / 2, j+1}+E_{z, i, j-1 / 2}+E_{z, i+1, j-1 / 2}\right) \\
&+\frac{\delta y}{8}\left[\left(\frac{\partial E_z}{\partial y}\right)_{i-1 / 2, j-1 / 4}-\left(\frac{\partial E_z}{\partial y}\right)_{i-1 / 2, j-3 / 4}\right] \\
&+\frac{\delta x}{8}\left[\left(\frac{\partial E_z}{\partial x}\right)_{i-1 / 4, j-1 / 2}-\left(\frac{\partial E_z}{\partial x}\right)_{i-3 / 4, j-1 / 2}\right] \,,  \label{eq:ct-1}
\end{align}
where the derivatived of EMF on each cell face is computed from the upwinded contact mode from the Riemann solver, i.e., suppressing the $j$ subscript,
\begin{equation}
\left(\frac{\partial E_z}{\partial y}\right)_{i-1 / 2}= 
\begin{cases}
\left(\partial E_z / \partial y\right)_{i-1}, & \text { for } v_{x, i-1 / 2}>0 \\ 
\left(\partial E_z / \partial y\right)_i, & \text { for } v_{x, i-1 / 2}<0 \\ 
\frac{1}{2}\left[\left(\frac{\partial E_z}{\partial y}\right)_{i-1}+\left(\frac{\partial E_z}{\partial y}\right)_i\right], & \text { otherwise },
\end{cases}
\end{equation}
The cell-centered derivatives of EMF are calculated from the face-centered EMF (Godunov flux) and cell-centered reference EMF $E^r_z$ (see Fig. \ref{fig:ct-2}):
\begin{equation}
\left(\frac{\partial E_z}{\partial y}\right)_{i, j-1 / 4}=2\left(\frac{E_{z, i, j}^r-E_{z, i, j-1 / 2}}{\delta y}\right), \label{eq:ct-2}
\end{equation}
For more details on how to compute the reference EMF in a certain timestep, see \cite{Stone:2008mh}. 
Since the NO is trained to predict the cell-centered volume-averaged magnetic field, from Eq. \ref{eq:ct-1} and \ref{eq:ct-2}, we replace the face-centered EMF from Riemann solver with that from NO prediction.

\end{document}